# SOLVING THE GRAPHENE ELECTRONICS CONUNDRUM: HIGH MOBILITY AND HIGH ON-OFF RATIO IN GRAPHENE NANOPATTERNED TRANSISTORS

Mircea Dragoman[1*], Adrian Dinescu[1], and Daniela Dragoman[2,3]

[1]National Institute for Research and Development in Microtechnology (IMT), P.O. Box 38-160, 023573 Bucharest, Romania,

[2]Univ. Bucharest, Physics Faculty, P.O. Box MG-11, 077125 Bucharest, Romania

[3]Academy of Romanian Scientists, Splaiul Independentei 54, 050094 Bucharest, Romania

**Abstract:**

Tens of graphene transistors with nanoperforated channels and different channel lengths were fabricated at the wafer scale. The nanoholes have a central diameter of 20 nm and a period of 100 nm, the lengths of the channel being of 1, 2, 4 or 8 μm. We have found that the mobility in these 2 μm-wide transistors varies from about 10400 cm$^2$/Vs for a channel length of 1 μm to about 550 cm$^2$/Vs for a channel length of 8 μm. Irrespective of the mobility value, in all transistors the on-off ratio is in the range $10^3$-$10^4$ at drain and gate voltages less than 2 V. The channel length-dependent mobility and conductance values indicate the onset of strong localization of charge carriers, whereas the high on-off ratio is due to bandgap opening by nanoperforations.

______________________________________________________________________

*Corresponding author: mircea.dragoman@imt.ro

## 1. Introduction

Graphene electronics, which was still a hot topic few years ago, is loosing interest at the expense of two-dimensional (2D) materials electronics based on transitional metal dichalcogenides (TMDs) and X-nes (phosphorene, silicene, germanene). An updated state-of-the-art can be found in [1]. Graphene electronics was almost abandoned because the key electronic device – the graphene field-effect transistor (FET), cannot be switched on and off due to the absence of a bandgap. The lack of bandgap in the graphene monolayer has as dramatic consequence the suppression of saturation and blocking regions, the graphene monolayer FET (GFET) working as a voltage controlled resistance, but not as a switch. However, the mobility in graphene FETs attains impressive values, for example 23600 $cm^2/Vs$ at room temperature in top-gated graphene nanoribbon FETs [2]. We have reported recently encapsulated GFETs with a nanoperforated channel, which show saturation and blocking regions tunable via the top gate voltage, and on/off ratios of at least $2\times10^3$ at room temperature at small drain and gate voltages, as well as a mobility of 2200 $cm^2/Vs$, higher than in many TMD monolayers such as $MoS_2$ or $WS_2$ [3]. Most TMDs show lower mobilities than Si, even as estimated from first principles [4], but excellent on/off ratios because they have a bandgap in the range of 1-2 eV. On the other hand, gapless X-nes, as graphene, show high mobilities, of 5200 $cm^2/Vs$, for example, in phosphorene FETs at room temperature [5], but are unstable in air, which is a huge drawback compared to graphene.

It is possible to solve the conundrum of graphene electronics: a GFET with a mobility having similar values as in graphene monolayer, or III-V semiconductor compounds based on GaAs or InP, and a high on-off ratio like in most 2D materials other than graphene or semiconductors? The issue is of huge practical importance, since a very high mobility is associated with a high-performance ultrafast transistor, i.e. an ultrafast switch. Such a switch could solve the main computer bottleneck, i.e. the stagnation of the clock speed of the computer CPU, which has not increased significantly since more than a decade.

We answer positively to the question before by showing that a GFET with a nanopatterned channel is in fact a high-mobility, high on-off ratio FET. Although the nanopatterning of graphene channels in GFETs was studied before, a careful perusing of the collection of papers mentioned in [3] and of the review [6] shows that previous results were based on unreliable lithographical methods, which lack on reproducibility. Therefore, no transconductance or mobility values in these structures were reported before, despite the fact that the first attempts to fabricate nanopatterned GFETs are seven years old. On the other hand, batch fabrication of nanopatterned FETs and their encapsulation [3] resulted in high transconductance, a mobility twice as large as in Si, and high on-off ratio. In this paper we extend our analysis of nanopatterned GFETs in [3] by scaling down their dimensions and performing a batch fabrication based on e-beam lithography of 90 nanopatterned GFETs with channel lengths of 1, 2, 4 and 8 µm. The aim is to study the dependence on the channel length of mobility, drain conductance and on-off ratio, and to elucidate the transport phenomena in these GFETs. We have found that (i) the mobility decreases from 10400 cm$^2$/Vs for the GFETs with 1 µm channel up to 550 cm$^2$/Vs for a channel length of 8 µm, indicating that strong localization of carriers takes place in GFETs, with an average localization length of 1.9 µm, and that (ii) the on-off ratio is higher than $10^3$ in all transistors at drain and gate voltages less than 2 V, reaching values up to $3\times10^4$ in some devices, due to the bandgap induced by nanopatterning.

## 2. Fabrication and characterization of nanopatterned GFETs

The graphene monolayer, grown by the CVD method, was transferred on a 4 inch Si/SO$_2$ substrate. The SiO$_2$ layer has a thickness of 300 nm and the Si substrate has high resistivity, greater than 8 kΩ·cm. Raman analysis was used to verify the graphene quality transfer and the persistence of graphene monolayer characteristic after certain critical fabrication processes involving especially PMMA removal and dielectric deposition. Raman analysis, performed on different areas of the wafer, show that the transferred CVD-grown graphene monolayers form in fact islands with maximum dimensions of 5 mm. These monolayer islands are surrounded and connected by a network of wrinkles, where all types of graphene, from bilayers up to multilayers, can be identified. In all tests, we considered that graphene monolayers preserve their properties if the D band is absent, if the G band has a peak around 1586 cm$^{-1}$ and the 2D peak is around 2640 cm$^{-1}$, and if the ratio between the 2D and G peaks is at least 2. The result was that we found mainly graphene monolayers in a large region, of about 40 percent of the wafer. Although the Raman map of the wafer is time consuming, it is necessary in order to make sure that the selected region has a smaller number of defects than the surrounding regions.

Further, we have cut the selected graphene chip from the wafer and started the fabrication process for 90 GFETs with nanopatterned channels. We have fabricated nanoperforated GFETs with channel lengths of $L$ = 1, 2, 4 and 8 μm. Twelve steps are necessary for batch fabrication of nanopatterned GFET. First, the graphene channels of all 90 GFETs are fabricated via e-beam patterning and reactive-ion etching, followed by nanoperforation by e-beam. The nanopatterning consists of a periodic array of holes, with a diameter of 20 nm and a period of at 100 nm, perforated in a graphene nanoribbon with a width $W$ = 2 μm. The holes near the nanoribbon boundaries have nearly 30 nm in diameter due to proximity effects in e-beam lithography. As will be seen in the following, this detrimental effect of nanolithography turns into an advantage for nanopatterned GFETs. The SEM image of a 2 µm-long nanopatterned GFET channel is displayed in Fig. 1. Then, source and drain contacts of Ti(10 nm)/Au(100 nm) are deposited via e-beam evaporation. Further, the negative electron resist HSQ, which plays the role of dielectric, is deposited with a thickness of 40 nm over the drain-source channel, and finally the Ti(10 nm)/Au(100 nm) gate electrode is deposited via e-beam evaporation. A SEM image of a nanoperforated GFET is displayed in Fig. 2(a), the more detailed SEM image in Fig. 2(b) showing the drain-source region, the HSQ dielectric layer and the top gate electrode. Additional information about the fabrication process is found in [3]. The difference in this case is that each fabrication step was carefully monitored using optical and electronic microscopy and Raman spectroscopy. In this way we discovered that about 30% of GFETs are not working due to the following reasons: (i) multiple cracks in the GFETs channel accompanied by resist debris (see Fig. 3(a)), and (ii) ripped graphene flakes from channel during processing (as illustrated in Fig. 3(b)). By far, the most encountered defect mechanisms are cracks in the channel, sometimes as large as the channel width, as a result of pre-existing cracks due to growth and transfer processes.

On-wafer electrical measurements of the nanopatterned GFETs were performed using a Keithley 4200 SCS equipped with low noise amplifiers at all outputs. The graphene chip is placed on the chuck of the probe station equipped with probe tips, the chuck being positioned inside a Faraday cage, which is closed during measurements for protection against external source noises. All measurements are performed at room temperature by grounding the source electrode. In order to verify the validity of our experimental data, we have repeated the measurements at different voltage steps and used dual-sweeps to find eventual hysteretic behaviours. We measured the drain current ($I_D$) versus drain voltage ($V_D$) at various gate voltages ($V_G$) and the drain current ($I_D$) versus gate voltage ($V_G$) at various drain voltages ($V_D$), finding no hysteresis or any dependence of the results at changing the sweeping rates. No smoothing procedures are introduced during the measurements or afterwards.

The main aim of collecting the data over working 60 nanopatterned GFETs is to calculate the conductance and mobility as a function of the channel length and to investigate the transport mechanism. For the range of channel lengths in our devices there are two possibilities: the mobility and conductance do not depend on *L*, behaviour consistent with a diffusive regime [7-8], or a dependence on *L* is observed for the two parameters, behaviour indicating that charge carrier wavefunction interference/phase effects are non-negligible.

We have calculated the mobility from experimental data using the formula:

$$\mu = g_m L / W C_g \ |V_D| = g_m L t / \varepsilon W \ |V_D| \tag{1}$$

where $g_m = \partial I_D / \partial V_G$ is the transconductance, *L* and *W* are the length and the width of the GFET channel, respectively, $C_g$ is the gate capacitance per unit area, *t* is the thickness of the gate dielectric, and $\varepsilon = \varepsilon_0 \varepsilon_r$, with $\varepsilon_0$ the vacuum permittivity and $\varepsilon_r$ the relative dielectric permittivity.

## 3. Results and discussions

For all devices, we have extracted the transconductance and drain conductance $g_D = \partial I_D / \partial V_D$ from the electrical characteristics of GFETs. Typical dependences of these parameters on $V_G$ for different drain voltages indicated in the inset and on $V_D$ for different gate voltages indicated in the inset are presented in Figs. 4(a) and 4(b), respectively, for a device with $L = 2$ μm. The Dirac point in this device, defined by $g_m = 0$ is found to be at around $V_G = 0$ V, the observed discontinuity in $g_D$ around $V_D = 0$ V being associated to a low-carrier transport regime in which any longitudinal asymmetry in the device becomes noticeable when the polarity of the drain voltage changes.

Because our main interest was to determine the dependence of the mobility on the channel length, we have calculated the maximum value of this parameter for GFETs with a given channel length using (1), and then displayed the results in Fig. 5. The room-temperature maximum mobility has impressive values, at least twice as large as in Si, for batch-fabricated GFETs with channel lengths up to 4 μm, the mobility for the device with $L = 1$ μm attaining the value of 10400 cm$^2$/Vs, which is a record for nanopatterned GFET. From Fig. 5 it follows that the mobility has an exponential dependence on the channel length, which suggests the onset of the strong localization phenomenon. A localization length associated to the carrier mobility cannot be extracted from Fig. 5 because the maximum mobilities were obtained at different gate and drain voltages: 0 V and 0.3 V for $L = 1$ μm, -0.5 V and 0.1 V at $L = 2$ μm, 0.5 V and -0.2 V for $L = 4$ μm and 8 μm.

In fact, the distinctive exponential dependence on $L$ that is associated to strong localization does not refer to mobility, but to the drain conductance. More precisely, in this transport regime

$$g_D(L) = g_{d0} \exp(-L/L_{loc}) \tag{2}$$

where $L_{loc}$ is the localization length. Such dependence is indeed found in the nanopatterned GFETs, as can be seen from Fig. 6(a) where we have represented the drain conductance as a function of the channel length for $V_G = 0$ V and at different drain voltages indicated in the inset. By fitting the experimental data with equation (2) we were then able to extract the localization length for different drain and gate voltages. The obtained results are shown in Fig. 6(b) for different $V_G$ values indicated in the inset. $L_{loc}$ varies between 1.25 µm and 2.5 µm at room temperature, with an average value of about 1.9 µm.

A first remark is that strong localization occurs whenever $L_{loc}$ is smaller than the phase coherence length, the exponential dependence of drain conductance on $L$ occurring as long as the channel length is in turn smaller than the phase coherence length. It follows thus that the latter parameter is at least about 8 µm at room temperature in a nanopatterned GFET. This is in itself a remarkable result. Indeed, in the vast majority of previous experiments on graphene antidot arrays the phase coherence length reached at most 400 nm at very low temperatures [9-11], so that only weak and not strong localization of charge carriers could be observed. However, a similar exponential decrease of conductance with $L$ as in our case and similar localization and phase coherence lengths have been observed in [7] at very low temperatures, of only 2 K. It is important to mention that the nanoholes in [7], with a diameter of 100 nm and a periodicity of 150 nm, patterned by e-beam on mechanically exfoliated graphene, have almost the same diameter, the channel being cut after nanopatterning.

In contrast, in our GFETs long room temperature localization and phase coherence lengths could be explained by the nonuniform nanohole diameters in the transverse direction, which induces charge carrier focusing/guiding along the middle part of the channel such that the charge carriers avoid the boundaries of the channels and the associated strong recombination centers. These non-uniformities in the hole diameters are induced by the proximity effects due to e-beam lithography. Indeed, we have estimated the bandgap induced by nanopatterning using the method described in [3] and have found that for a nanohole array with a period of 100 nm and a diameter of 20 nm the corresponding bandgap is about 0.16 eV, but this parameter increases to 0.2 eV if the nanohole diameter becomes 30 nm. As a result, the charge carriers are guided through the central part of the channel and recombinations at channel boundaries are strongly reduced. It should be mentioned also that any disorder in the fabrication of the nanohole array is predicted to favor the onset of strong localization [12], a strong disorder in low-mobility graphene films being found to preserve weak localization up to room temperatures [13].

The fact that nanopatterning induces a bandgap is expected to enhance the on/off ratio of GFETs. More precisely, for a bandgap of $E_g$ = 0.16 eV, we obtain $\exp(E_g/k_BT) \cong 500$, while for $E_g$ = 0.16 eV, the corresponding term is $\exp(E_g/k_BT) \cong 2200$. Thus, we expect an on/off ratio of about 1000 assuming that charge carriers pass through regions with both smaller and larger nanoholes. Indeed, from Fig. 7(a), which shows the $I_D - V_G$ dependence of a GFET with $L$ = 2 μm at different drain voltages indicated in the inset, it follows that the on/off ratio is higher than $10^3$ for all drain and gate voltages. To check if this result is valid for GFETs with different lengths, we have represented in Fig. 7(b) the on/off ratio between drain currents at $V_D = 0$ V and 2 V as a function of the channel length for several gate voltages indicated in the inset. In all cases this ratio is larger than $1.6 \times 10^3$, attaining the value of $3 \times 10^4$ for the GFET with $L$ = 2 μm at low $V_G$ values. This excellent result for the device with a 2 μm-long channel is due to the fact that the off current is lower than in other GFETs.

## 4. Conclusions

We have found that the mobility in nanopatterned GFETs in the range 10400 cm$^2$/Vs for a channel length of 1 μm up to 550 cm$^2$/Vs for a channel length of 8 μm., all transistors displaying a high on-off ratio is in the range $10^3$-$10^4$. All 60 GFETs were measured and the dispersion of data among different transistors regarding mobility or on/off ratio was around 11-12%. These unexpected results are explained by the fact that the GFET is a nanopatterned non-uniform array and the current is flowing only through the middle of channel being suppressed towards the channel boundaries avoiding in this way edge-roughness scattering effects. The non-uniformity of nanopatterned graphene channel are introduced by the proximity effects originating from e-beam lithography, but can be produced also systematically by careful design of the channel. Batch fabrication of tens of nanopatterned GFETs having high-performances described above are reopening the nearly forgot hope that graphene FETs could be used in digital applications replacing silicon and in high-frequencies low noise deices up to THz frequencies. The single obstacle in graphene electronics remain the quality of graphene grown by CVD at the wafer level which was continuously improved in the last years now 6 inch wafers being available.

**References**

[1] M. Dragoman and D. Dragoman, *2D Nanoelectronics. Physics and Devices of Atomically Thin Materials*, Springer (2017).

[2] L. Liao, J. Baib, Y. Qua, Y.-C. Linb, Y. Lib, Yu Huang, and X. Duan, High-κ oxide nanoribbons as gate dielectrics for high mobility top-gated graphene transistors, PNAS 107, 6711-6715 (2010).

[3] M. Dragoman, A. Dinescu, and D. Dragoman, Room temperature nanostructured graphene transistor with high on/off ratio, Nanotechnology 28, 015201 (2017).

[4] W. Zhang, Z. Huang, W. Zhang, and Y. Li, Two-dimensional semiconductors with possible high room temperature mobility, Nano Research 7, 1731-1737 (2014).

[5] G. Long, D. Maryenko, J. Shen, S. Xu, J. Hou, Z. Wu, W.K. Wong, T. Han, J. Lin, Y. Cai, R. Lortz, and N. Wang, Achieving ultrahigh carrier mobility in two-dimensional hole gas of black phosphorous, Nano Letters 16, 7768-7771 (2016).

[6] J. Yang, M. Ma, L. Li, Y. Zhang, W. Huang, and X. Dong, Graphene nanomesh: new versatile materials, Nanoscale 6, 13301-12313 (2014).

[7] H. Zhang, J. Lu, W. Shi, Z. Wang, T. Zhang, M. Sun, Y. Zheng, Q. Chen, N. Wang, J.-J. Lin, and P. Sheng, Large-scale mesoscopic transport in nanostructured grapheme, Phys. Rev. Lett. 110, 066805 (2013).

[8] A. Venugopal, J. Chan, X. li, C.W. Magnuson, W.P. Kirk, L. Colombo, R.S. Ruoff, and E.M. Vogel, Effective mobility of single-layer graphene transistors as a function of channel dimensions, J. Appl. Phys. 109, 104511 (2011).

[9] T. Shen, Y.Q. Wu, M.A. Capano, L.P. Rokhinson, L.W. Engel, and P.D. Ye, Magneto-conductance oscillations in graphene antidot arrays, Appl. Phys. Lett. 93, 122102 (2008).

[10] J. Eroms and D. Weiss, Weak localization and transport gap in graphene antidot lattices, New J. Phys. 11, 095021 (2009).

[11] A. Sander, T. Preis, C. Schell, P. Giudici, K. Watanabe, T. Taniguchi, D. Weiss, and J. Eroms, Ballistic transport in graphene antidot lattices, Nano Letters 15, 8402-8406 (2015).

[12] Z. Fan, A. Uppstu, and A. Harju, Electronic and transport properties in geometrically disordered graphene antidot lattices, Phys. Rev. B 91, 125434 (2015).

[13] J. Han, S. Wang, D. Qian, F. Song, X. Wang, X. Wang, B. Wang, M. Han, J. Zhou, Room-temperature observations of the weak localization in low-mobility graphene films, J. Appl. Phys. 114, 214502 (2013).

**Figure captions**

Fig. 1 SEM image of a nanopatterned GFET channel with $L = 2$ μm.

Fig. 2 SEM images of (a) the nanopatterned GFET, and (b) a corresponding detail of the source-drain region.

Fig. 3 Defects in a GFET channel: (a) multiple cracks accompanied by resist debris, and (b) ripped graphene flakes.

Fig. 4 (a) Dependences of transconductance on $V_G$ for different drain voltages indicated in the inset and (b) dependence of drain conductance on $V_D$ for different gate voltages indicated in the inset for a GFET with $L = 2$ μm

Fig. 5 Dependence of the maximum mobility on the GFET channel length

Fig. 6 (a) Dependence of the drain conductance on the channel length for $V_G = 0$ V and at different drain voltages, and (b) the extracted localization length as a function of the drain voltage for different $V_G$ values indicated in the inset.

Fig. 7 (a) $I_D - V_G$ dependence of a GFET with $L = 2$ μm at different drain voltages indicated in the inset, and (b) the on/off ratio between drain currents at $V_D = 0$ V and 2 V as a function of the channel length for several gate voltages indicated in the inset.

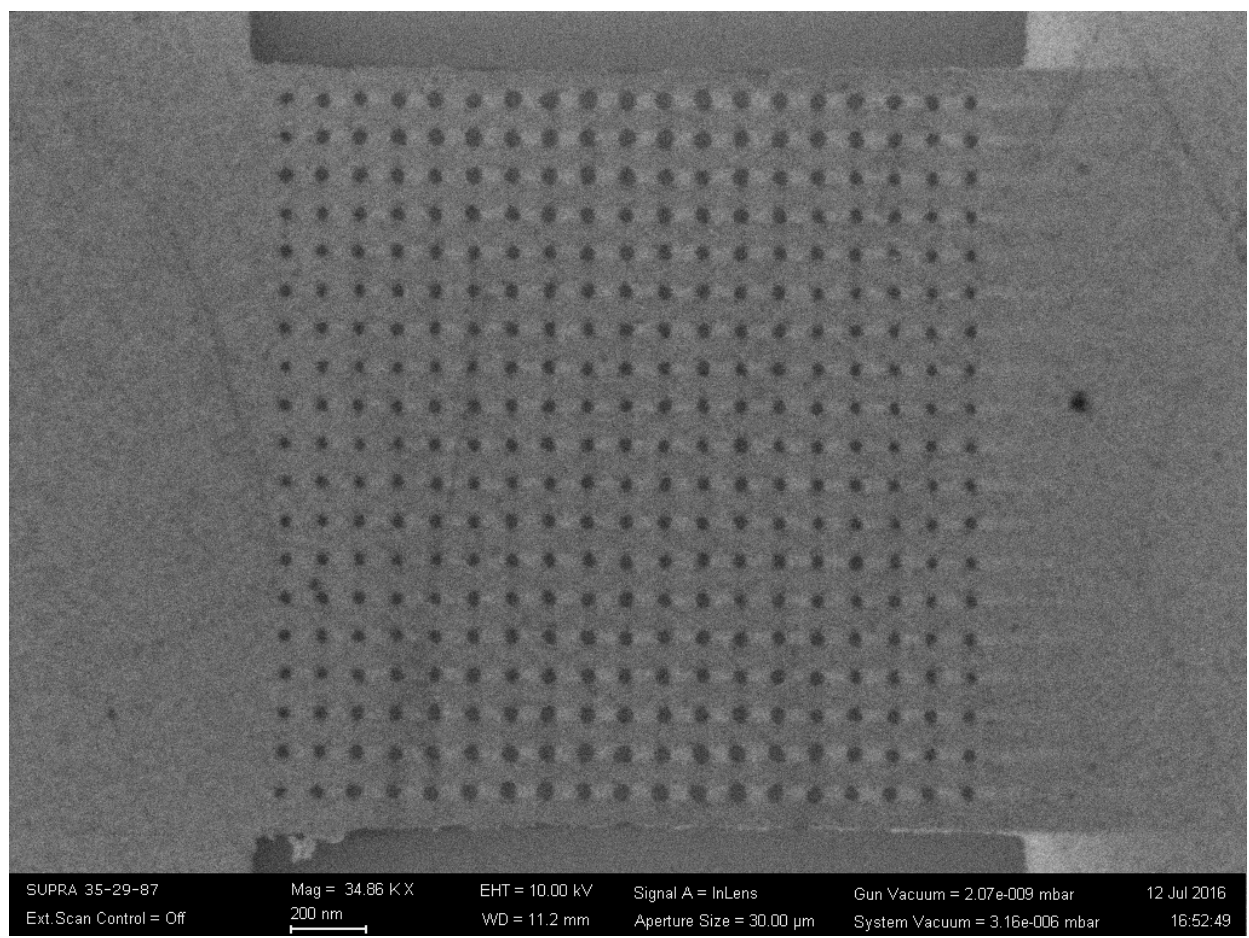

Fig. 1

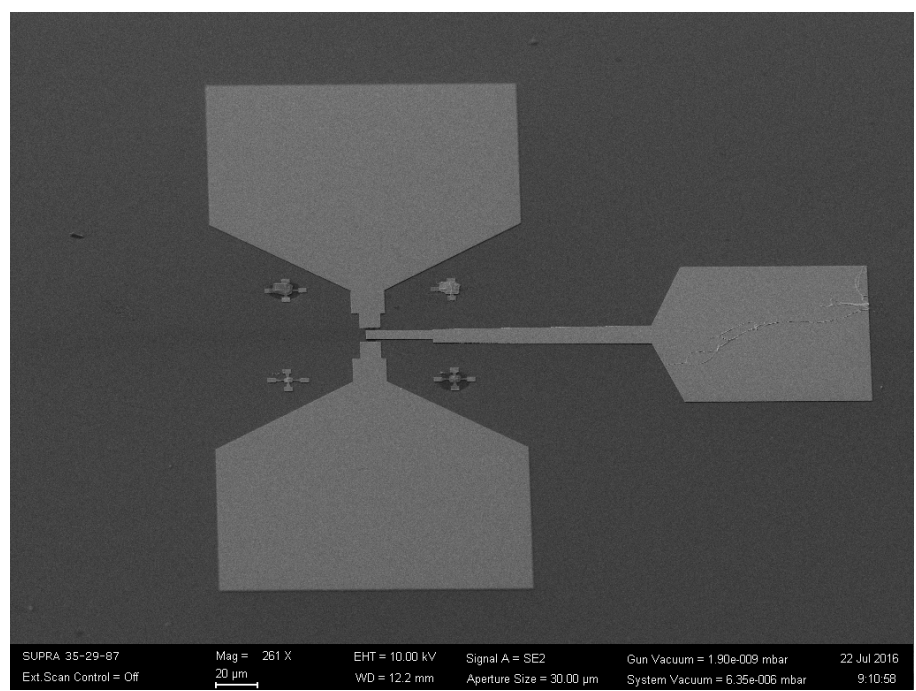

(a)

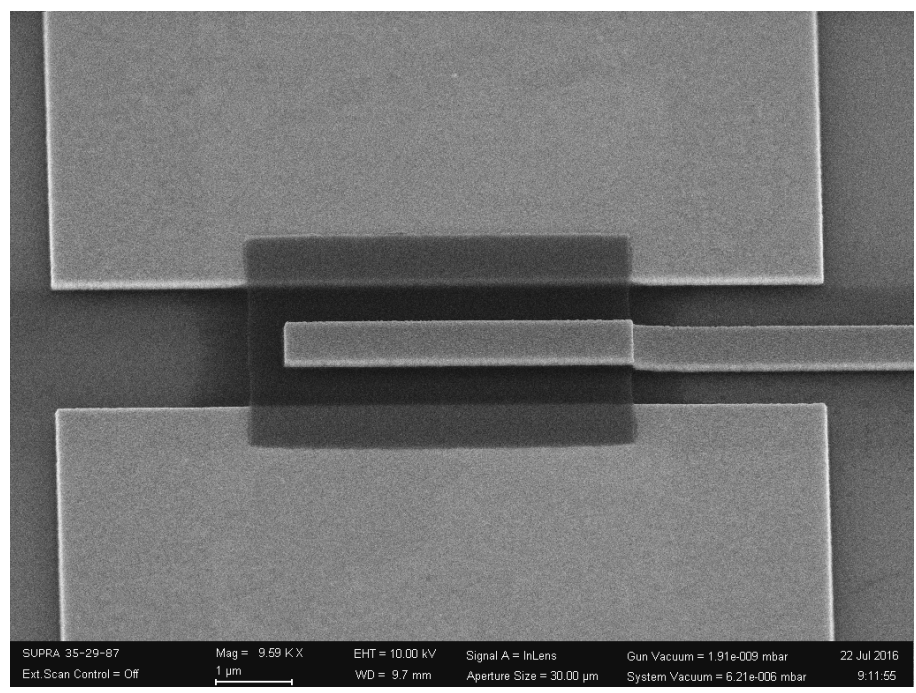

(b)

Fig. 2

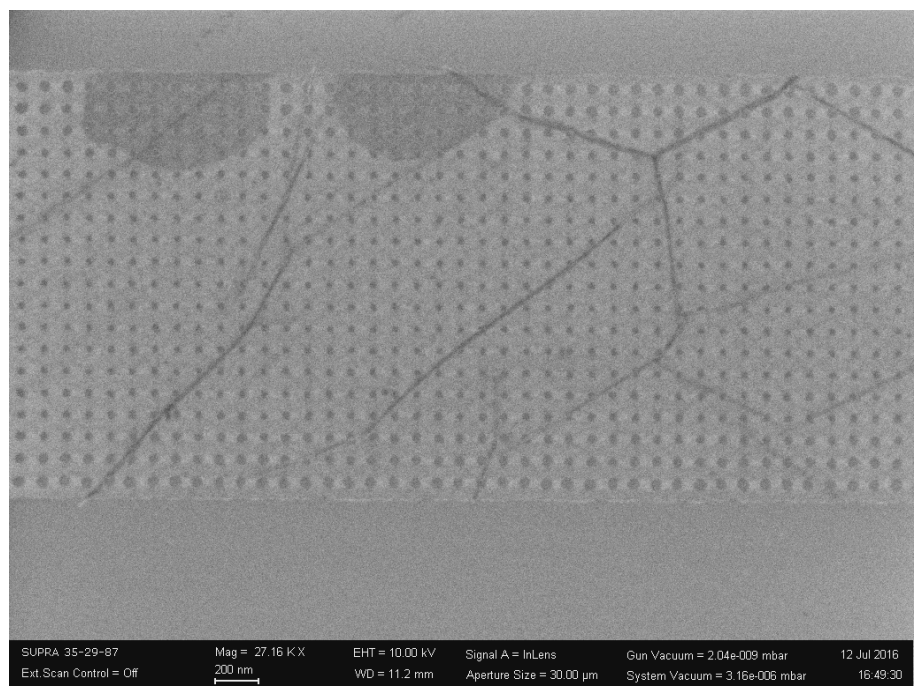

(a)

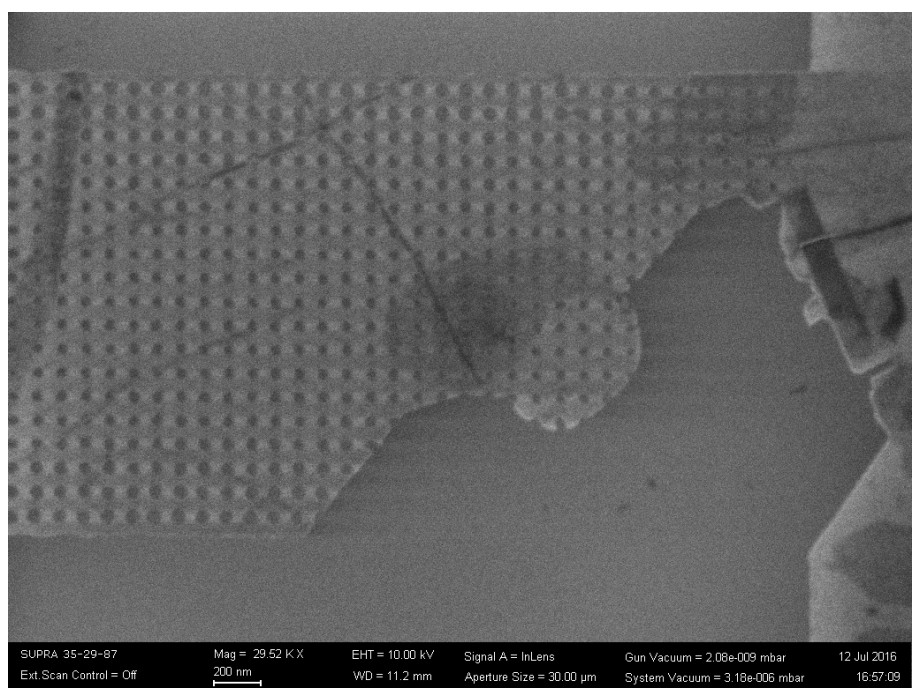

(b)

Fig. 3

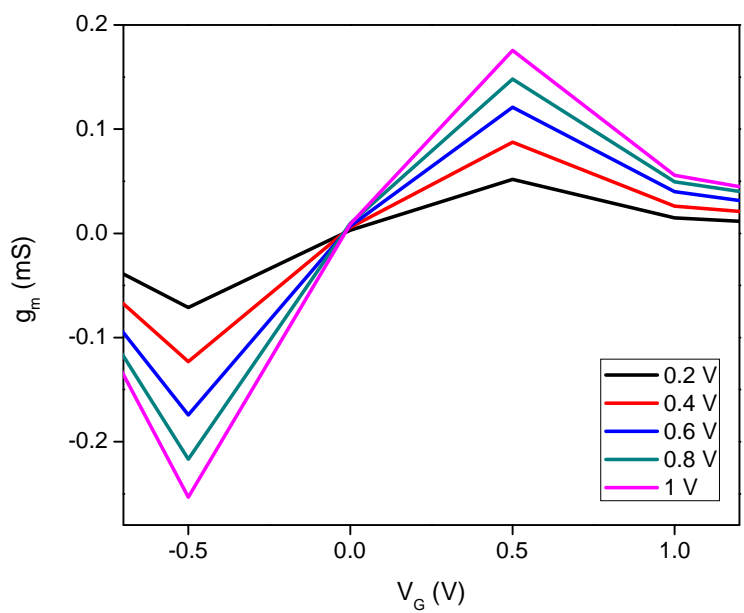

(a)

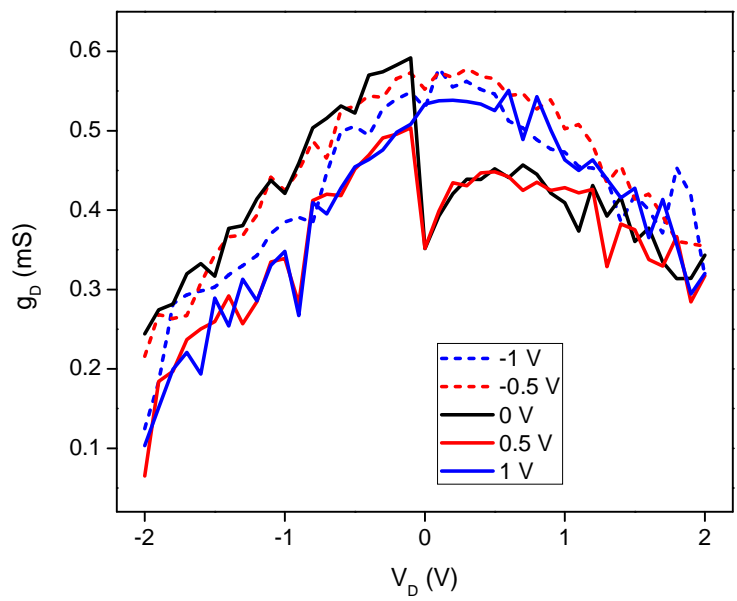

(b)

Fig. 4

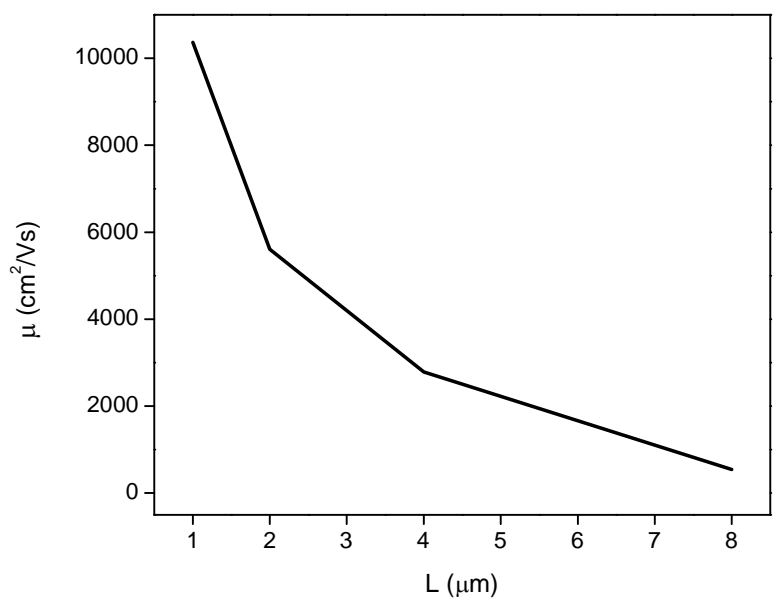

Fig. 5

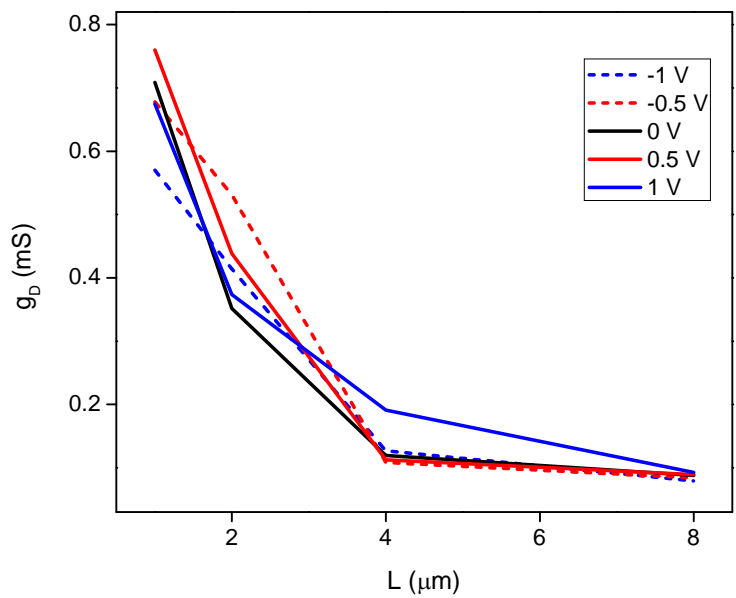

(a)

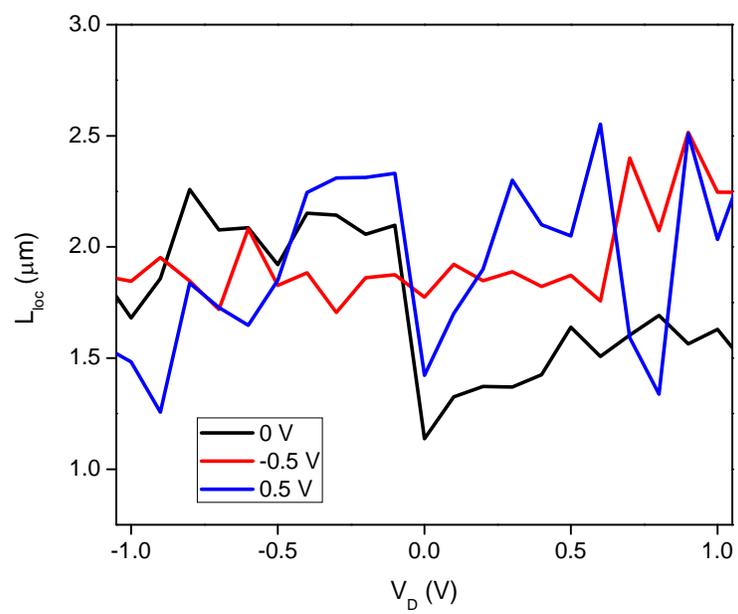

(b)

Fig. 6

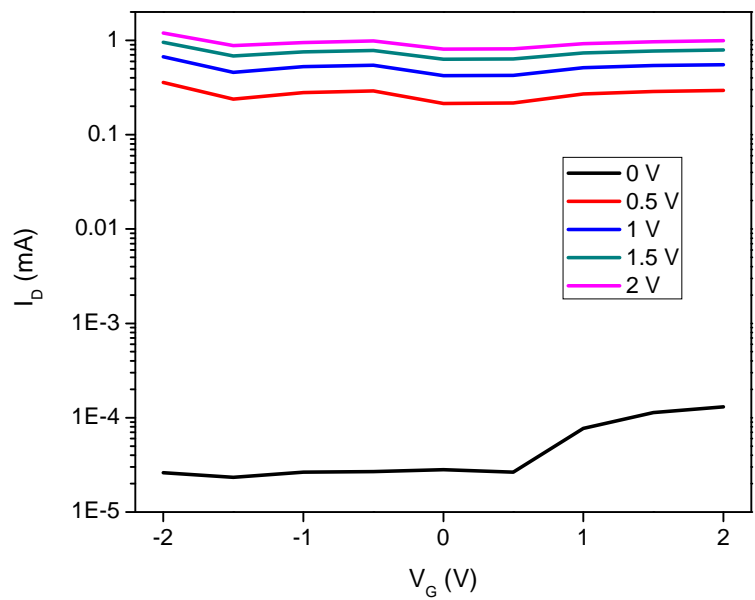

(a)

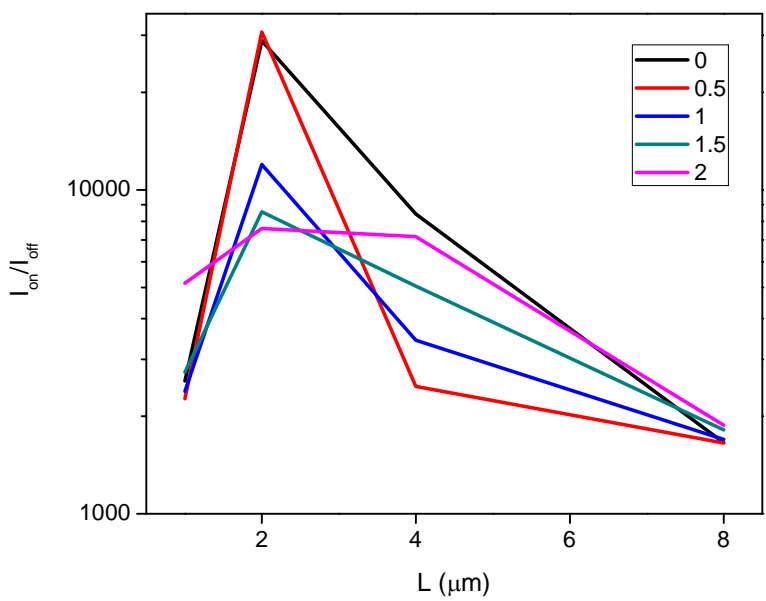

(b)

Fig. 7